\documentclass[prb,amsmath,amssymb,reprint]{revtex4-2}
\usepackage{amsmath}
\usepackage{graphicx}
\usepackage{bm}
\usepackage{xr}
\usepackage{todonotes}
\usepackage{enumitem}
\usepackage{float}
\begin{document}

\newcommand{\red}[1]{\textcolor{red}{#1}}
\newcommand{\blue}[1]{\textcolor{blue}{#1}}

\title{$\Delta$-model correction of Foundation Model based on the models own understanding}
\author{Mads-Peter Verner Christiansen}
\author{Bjørk Hammer}
    \email{hammer@phys.au.dk}
\affiliation{Center for Interstellar Catalysis, Department of Physics and Astronomy, 
    Aarhus University, DK‐8000 Aarhus C, Denmak}

\date{\today}

\begin{abstract}
Foundation models of interatomic potentials, so called universal potentials, may require fine-tuning
or residual corrections when applied to specific subclasses of
materials. In the present work, we demonstrate how such augmentation
can be accomplished via $\Delta$-learning based on the representation
already embedded in the universal potentials. The $\Delta$-model introduced is a
Gaussian Process Regression (GPR) model and various types of aggregation
(global, species-separated, and atomic) of the representation vector
are discussed. Employing a
specific universal potential, CHGNet [Deng \textit{et al.},
Nat.\ Mach.\ Intell.\ \textbf{5}, 1031 (2023)], in a global structure optimization setting,
we find that it correctly describes the energetics of the "8" Cu oxide, which is an ultra-thin oxide film on Cu(111).
The universal potential model
even predicts a more favorable structure compared to that
discussed in recent DFT-based literature. Moving to sulfur adatom
overlayers on Cu(111), Ag(111), and Au(111) the CHGNet model,
however, requires corrections. We demonstrate that these are
efficiently provided via the GPR-based $\Delta$-model formulated on the CHGNet's own
internal atomic embedding representation. The need for corrections
is tracked to the scarcity of metal-sulfur atomic environments in
the materials project database that CHGNet is trained on leading to an overreliance on 
sulfur-sulfur atomic environments. Other
universal potentials trained on the same data, MACE-MP0, SevenNet-0,
and ORB-v2-only-MPtrj show similar behavior, but with varying
degrees of error, demonstrating the general need for augmentation
schemes for universal potential models.
\end{abstract}

\maketitle

\section{Introduction}

Quantum mechanical calculations of material properties have been a corner stone 
of material science for the past several decades. This encompasses e.g. the 
prediction or theoretical explanation of phase diagrams of solids or surfaces. 
In recent years the advancement of machine learning (ML) techniques as a set of tools to 
reduce the computational expense of such studies has become widespread \cite{wangMachineLearningInteratomic2024}.

Perhaps the most common use of ML is that of replacing expensive first-principles 
calculations of total energies with orders of magnitude faster machine-learning interatomic 
potentials (MLIPs). Early works in this area included Behler-Parinello neural networks 
and the Gaussian Approximation Potentials of Bartok et. al \cite{Behler2007,Bartok2010}. 
Since then many advances have been made improving the accuracy and data efficiency of these potentials \cite{ruppFastAccurateModeling2012, behlerConstructingHighdimensionalNeural2015, 
botuMachineLearningForce2017, gilmerNeuralMessagePassing2017, schuttSchNetContinuousfilterConvolutional2017,
thomasTensorFieldNetworks2018, andersonCormorantCovariantMolecular2019, unkePhysNetNeuralNetwork2019, 
schuttEquivariantMessagePassing2021, deringerGaussianProcessRegression2021, batznerEquivariantGraphNeural2022, 
montes-camposDifferentiableNeuralNetworkForce2022, batatiaMACEHigherOrder2023, anstineMachineLearningInteratomic2023}. 
With these improvements many tasks in computational material science have benefitted from 
the efficiency they offer, however often demanding the construction of a task 
specific dataset, thus still requiring expensive ab-initio calculations. 

Recently datasets containing ab-initio properties of atomic configurations covering 
vast regions of chemical space have become available \cite{dengCHGNetPretrainedUniversal2023,chanussotOpenCatalyst20202021,tranOpenCatalyst20222023}.
These datasets have enabled the development of broadly applicable atomistic models—universal potentials—some of which share characteristics 
with foundation models in their transferability and adaptability. 
One such dataset is the Materials Project Trajectory (MPtrj) containing 
among others energies and forces for some $\sim$1.6 million structures extracted 
from the Materials Project. Several MLIPs have been trained on this dataset including 
CHGNet, MACE-MP0, SevenNet-0, and ORB-v2-only-MPtrj \cite{dengCHGNetPretrainedUniversal2023, batatia2023foundation, kim_sevennet_mf_2024,neumannOrbFastScalable2024}. 

One topic that will benefit from these potentials is the determination of phase stability under given
thermodynamic conditions, a problem often addressed in computational materials science for catalysis.
One strategy in this domain is global structure optimization, finding the most 
energetically, in terms of total energy, favorable geometries of an atomic system followed by a 
thermodynamic analysis to identify the phases with the lowest Gibbs free energy. 
The optimization step requires exploration of the potential energy surface, a 
task that has received much research interest in order to provide effective algorithms \cite{tongAcceleratingCALYPSOStructure2018, Todorovic2019, Bisbo2020, timmermannMathrmIrOSurface2020,yangMachinelearningAcceleratedGeometry2021, christiansenAtomisticGlobalOptimization2022, chenAutomatedSearchOptimal2022, larsenMachinelearningenabledOptimizationAtomic2023,wangMAGUSMachineLearning2023, lyngbyBayesianOptimizationAtomic2024,wanzenbockClinamen2FunctionalstyleEvolutionary2024,noordhoekAcceleratingPredictionInorganic2024,luoDeepLearningGenerative2024,pitfieldPredictingPhaseStability2024,cerasoliEffectiveOptimizationAtomic2024, hessmannAcceleratingCrystalStructure2025}

The construction of accurate phase diagrams depends critically on the
accuracy of the underlying total energy description. In our recent
work on describing
the global optimal structure of silicate clusters and ultra-thin oxide
films on Ag(111) with CHGNet, it was found necessary to augment
the CHGNet model with $\Delta$-learning in order to get the correct order
of stability of low-lying structures \cite{pitfieldAugmentationUniversalPotentials2025}. In that work, the $\Delta$-model
was built on a representation involving the SOAP descriptor for each
atom. In the present work, we introduce the necessary formalism for
eliminating the need of such a descriptor and instead use the internal
representation of the atoms in the CHGNet when constructing the $\Delta$-model.
Employing the method in a global optimization setting, we find
the resulting corrections for ultra-thin oxide films on
Cu(111)-$c(8\times 4)$ to be small. The search results in an oxide
film structure that represents a reinterpretation of the
experimentally found structure. This result clearly testifies to 
CHGNets ability as a universal potential.

In contrast, for sulfur ad-atom layers on Cu(111), Ag(111),
and Au(111) we find that the correction terms resulting from the delta
model are more critical, in particular for Ag(111) and Au(111). By
analyzing the similarity of local atomic environments in the sulfur
ad-layer systems and in the training database for CHGNet, we trace the
origin of the low accuracy in these systems to the lack of relevant
Ag-S and Au-S local environments in the training data.

The paper is structured as follows: First we introduce the
methodologies employed, this includes an account of the descriptor
we employ which has been extracted from CHGNet, the Gaussian Process
Regression scheme that we use, and a summary of the global optimization
(GO) algorithm used for structural searches. Next, we present the
computed phase diagrams for O/Cu(111) and S/Au(111), showing the need
for corrections in the latter case. The paper proceeds by comparing
the CHGNet behavior for a specific S-coverage on Cu(111), Ag(111), and
Au(111) and relates that to the occurrence of relevant structures in
the MPtrj dataset thereby shedding light on why CHGNet gets the
relatively simple sulfur overlayer structure so wrong when it is
capable of predicting a previously undiscovered phase of the copper
oxide. The paper ends by identifying that the issues for the sulfur
ad-layers pertains to other foundation models, MACE-MP0, SevenNet-0,
and ORB-v2-only-MPtrj, that have all been trained on the same dataset as CHGNet.

\section{Methodology}

\subsection{Description of Atomic Environments}

\begin{figure}
    \centering
    \includegraphics[width=0.40 \textwidth]{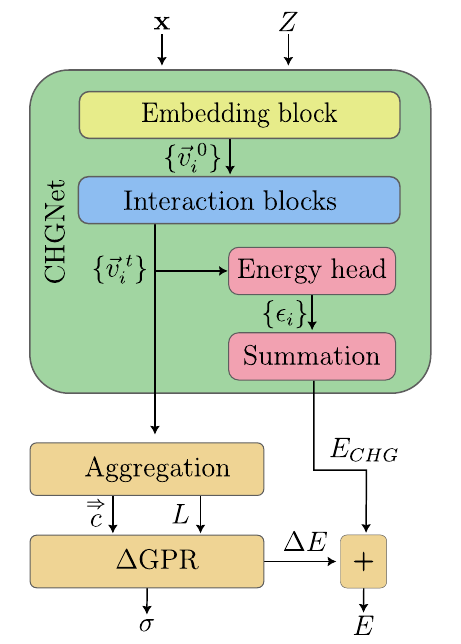}
    \caption{Illustration of our use of CHGNet, including a summarized version of CHGNet and our additional $\Delta$-model 
    using the CHGNet atomic representations.}
    \label{fig:architecture}
\end{figure}

The development and success of machine learning tools in computational materials science 
have been driven by improvements to the description of atomic environments. The field 
initally relied on handcrafted descriptors, such as Behler-Parinello symmetry functions \cite{behlerAtomcenteredSymmetryFunctions2011}, 
the Valle-Oganov fingerprint \cite{oganovHowQuantifyEnergy2009} or the Smooth Overlap of Atomic Positions (SOAP) formalism \cite{bartokRepresentingChemicalEnvironments2013}.
Since then neural networks such as SchNet that are capable of learning useful representations directly from 
Cartesian coordinates have become widespread \cite{schuttSchNetContinuousfilterConvolutional2017, pmlr-v139-schutt21a,batznerE3equivariantGraphNeural2022}.

With advances in neural network architecture and the emergence of large datasets a multitude of foundation 
models have been introduced. One such foundation model is CHGNet, which has been trained on the MPtrj dataset \cite{dengCHGNetPretrainedUniversal2023}. 
CHGNet is a  graph neural network where message-passing operations are used to iteratively update 
the representation of each atom with information from its surrounding neighbors.
The layout of CHGNet is shown schematically in the green box of Fig.\ \ref{fig:architecture}.
A prediction from CHGNet, and likewise for the majority of other graph based MLIPs, 
is calculated by first updating the atomic representations with message-passing 
and then passing these refined descriptors to a prediction head, that transforms and 
aggregates the descriptors in order to produce a prediction. Typically, an energy prediction 
is made by transforming the high-dimensional descriptors to a scalar for each atom and summing these together. 
We will use $\vec{v_i}$ to denote representations of atomic environments extracted from CHGNet. These 
are 64 dimensional vectors one for each atom in an atomic configuration. These descriptors may be extracted 
and used for other tasks, such as $\Delta$-learning as shown schematically with the brown boxes in Fig.\ \ref{fig:architecture}.

\begin{figure*}[ht!]
    \centering
    \includegraphics[width=0.8\textwidth]{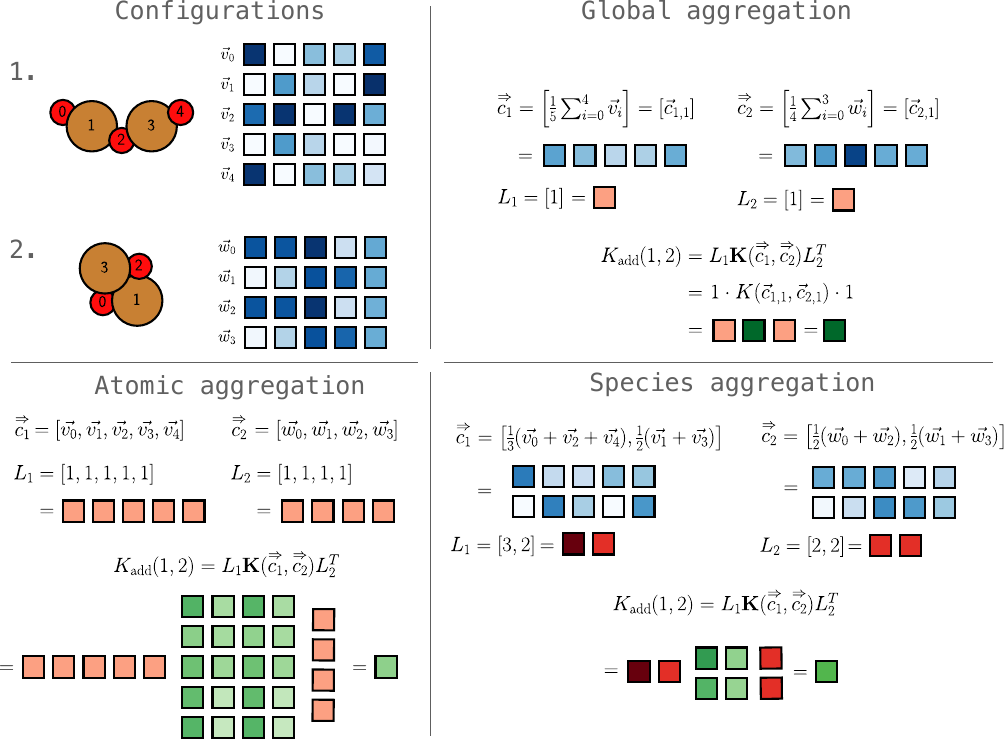}
    \caption{Illustration of aggregation procedures for two configurations. For \textit{global aggregation} each 
    configuration is represented by a single vector and the additive kernel reduces to a simple kernel between 
    the vectors for each configuration. For \textit{atomic aggregation} each configuration is represented by 
    vectors describing the environment of each atom with unit weight $L$. Finally for \textit{species aggregation} each 
    configuration is represented by two vectors each describing the average oxygen and copper, as the first configuration 
    has more oxygen than the second they have different $L$.}
    \label{fig:gpr_explain}
\end{figure*}

For a full description of the architecture of CHGNet, we refer to
Figure 1 of the original publication Ref.\ 
\cite{dengCHGNetPretrainedUniversal2023}. For the present purposes, it
serves to summarize it as:
\begin{enumerate}
    \item Initialize relevant properties for the message-passing, including initial descriptors, $\vec{v_i}^0$.
    \item Apply message-passing interaction blocks $t-1$ times resulting in descriptors $\vec{v_i}^{t-1}$ for each atom $i$. 
    \item Predict magnetic moments using vectors $\vec{v_i}^{t-1}$.
    \item Apply one more message-passing block to compute vectors $\vec{v_i}^{t}$.
    \item Predict atomic energies $\epsilon_i$ using the vectors $\vec{v_i}^{t}$.
    \item Sum local energies to predict total energy $E = \sum_i \epsilon_i$.
\end{enumerate}
CHGNet consists of a representation part, that uses message-passing 
to construct descriptors of atomic environments, steps 1 through 4, and a fully-connected neural network 
that converts each atomic descriptor to an atomic energy and the atomic energies 
are summed to compute the total energy, steps 5 and 6. The individual atomic descriptors are 64-dimensional 
vectors. We use the final atomic descriptors $\vec{v_i}^{t}$ as the basis for a $\Delta$-model described in the next section.

We are interested in using CHGNet for systems that are quite dissimilar to the the MPtrj training set, 
specifically surfaces. In order for the $\Delta$-model described in the next section to be a viable strategy, 
the descriptors produced by CHGNet need to be amendable to learning. 
While graph convolutions provide a mathematically rich framework for learning representations, 
they also introduce implicit biases that guide the representations, 
making the learned descriptors useful even beyond the training dataset.
We find this to be the case, as evidenced by the results we present in later sections.

\subsection{Additive Gaussian Process Regression}

We follow our recent proposal \cite{pitfieldAugmentationUniversalPotentials2025} 
and employ a Gaussian Process Regression (GPR) model in a $\Delta$-learning context for correcting 
the universal potential in regions where it makes incorrect predictions. However, 
in this work, we introduce a different formalism that ties more closely together with the neural network.

Customarily the predicted mean of a GPR is given by 

\begin{equation}
m(R) = k(R, \mathbf{X})[K(\mathbf{X},\mathbf{X})+\sigma_n^2\mathbf{I}]^{-1} y.
\label{eq:regular_gpr_mean}
\end{equation}
Where $R$ is the representation of a query object, $\mathbf{X}$ are representations of the 
training data, typically a matrix where each row is a feature vector but can more generally
be considered as a set of representations of the training examples. Each training example comes 
with a corresponding target $y$ and $\sigma^2_n$ is the variance of the assumed noise and $\mathbf{I}$ is the identity matrix. 
The representation $R$ may describe the full object through descriptors of parts of the object. 
To facilitate this, an additive kernel may be defined:
\begin{equation}
    K_{\mathrm{add}}(R_v, R_w) = \sum_i^{n_v}\sum_j^{n_w} L^v_i L^w_j K(\vec{v_i}, \vec{w_j}).
\end{equation}
Here $R_v$, the representation of object $v$, which consists of $(\stackrel{\Rightarrow}{c_v}, L_v)$ where 
$\stackrel{\Rightarrow}{c_v} = \left[\vec{v}_1, \vec{v}_2, .., \vec{v}_{n_v}\right]$ is a collection of vectors describing $v$ and $L^v_i$ is the 
number of contributions associated with each vector $\vec{v_i}$ -- likewise $R_w = (\stackrel{\Rightarrow}{c_w}, L_w)$. 
This type of kernel can compare two objects represented by a different number of vectors 
and with a different number of contributions. If $\stackrel{\Rightarrow}{c_v} = [\vec{v}_1]$, $\stackrel{\Rightarrow}{c_w} = [\vec{w}_1]$ and $L^v_1 = L^w_1 = 1$ 
it is evident that the original kernel $K(\vec{v_1}, \vec{w_1})$ is recovered. 
This introduces the additional property $L$, which we may also use to write the additive kernel in matrix form
\begin{equation}
    K_{\mathrm{add}}(R_v, R_w) = L_v K(\stackrel{\Rightarrow}{c_v}, \stackrel{\Rightarrow}{c_w}) L_w^T
\end{equation}
With this expression we can write Eq. \eqref{eq:regular_gpr_mean} with an additive kernel as,

\begin{equation}
    m(R_v) = L_v k(\stackrel{\Rightarrow}{c_v}, \mathbf{X}) L_X^T [L_X K(\mathbf{X}, \mathbf{X}) L_X^T + \sigma_n^2 \mathbf{I}]^{-1} y.
\end{equation}

For the task of learning total energies of atomic configurations starting from vectors describing the 
environment of individual atoms this leaves us with several options of how to construct the representations $R$. 
This is essentially a choice of an aggregation procedure
\begin{itemize}[itemsep=0pt]
    \item Global aggregation: $\stackrel{\Rightarrow}{c} = [\frac{1}{N}\sum_i^N \vec{v_i}$] and $L = [N]$ or $L=[1]$.
    \item Atomic aggregation: $\stackrel{\Rightarrow}{c} = [\vec{v_1}, .., \vec{v_N}]$ and $L = [1, .., 1]$
    \item Species aggregation: $\stackrel{\Rightarrow}{c} = [\vec{c_1}, \vec{c_2}, .., \vec{c_M}]$ where $\vec{c_m} = \frac{1}{N_m}\sum_{i}^N \vec{v_i} \cdot \delta(\mathcal{Z}_m, Z_i)$ and $L = [N_1, N_2, .., N_M] $
\end{itemize}
Where $N$ is the total number of atoms in a configuration, $\vec{v_i}$ are atomic descriptors, $M$ is the number of 
different species, $N_m$ is the number of atoms of species $\mathcal{Z}_m$, $Z_i$ is the species of atom $i$ and $\delta$ is the Kronecker delta function.
These aggregation procedures are depicted in Figure \ref{fig:gpr_explain}.
With a global aggregation scheme the model learns the total property directly and by
using $L=[N]$ it is capable of learning from data involving different amounts of atoms (unlike the situation if
$L=[1]$ is chosen, as is often done).
With 
atomic aggregation the model learns atomic energies such that they sum to the total energy. 
Similarly, with species aggregation the model learns the average energy of each species 
which in a sum weighted by the number of atoms of each species yields the total energy.
Global aggregation amounts to attributing one feature vector to the entire configuration in which 
case Eq. \eqref{eq:regular_gpr_mean} may be used. Atomic aggregation is equivalent to popular 
techniques such as GAP, but introduces poor computational scaling as the number of atomic 
environments in a training set may be very large -- generally necessitating the introduction of 
approximate GPR techniques such as the use of a sparsified GPR, that uses a subset of the data as inducing points to make 
training feasible and limit prediction time. Finally using \textit{species aggregation} a configuration is
described by as many vectors as there are unique atomic species with each vector describing the average environment of 
that species and $l_m$ counting the number of atoms of each species. This reduces the computational 
expense and eliminates the need for approximate GPR techniques while offering improved 
resolution compared to global aggregation. A similar scheme has previously been used for filtering 
of atomic structures \cite{kovacsLoGANLocalGenerative2024}.

We use this species aggregation GPR model in combination with CHGNet in a $\Delta$-learning 
scheme, where the GPR learns to correct the errors of CHGNet, we will refer to this 
as $\Delta$GP-CHGNet for the remainder of the article.
A final note on the GPR model, in addition to extracting features from CHGNet we 
may also leverage the automatic differentiation capabilities of PyTorch that 
CHGNet is written in. This means derivatives of the total $\Delta$GP-CHGNet energy can be computed at 
essentially no additional expense compared to those of just CHGNet. Additionally, 
this way of evaluating forces does not require the implementation of any analytical 
derivatives, be it of features or kernels, which is normally the most challenging 
and error-prone part of implementing a GPR for the prediction of atomistic properties.

\begin{figure*}[ht!]
    \centering
    \includegraphics[width=0.7\textwidth]{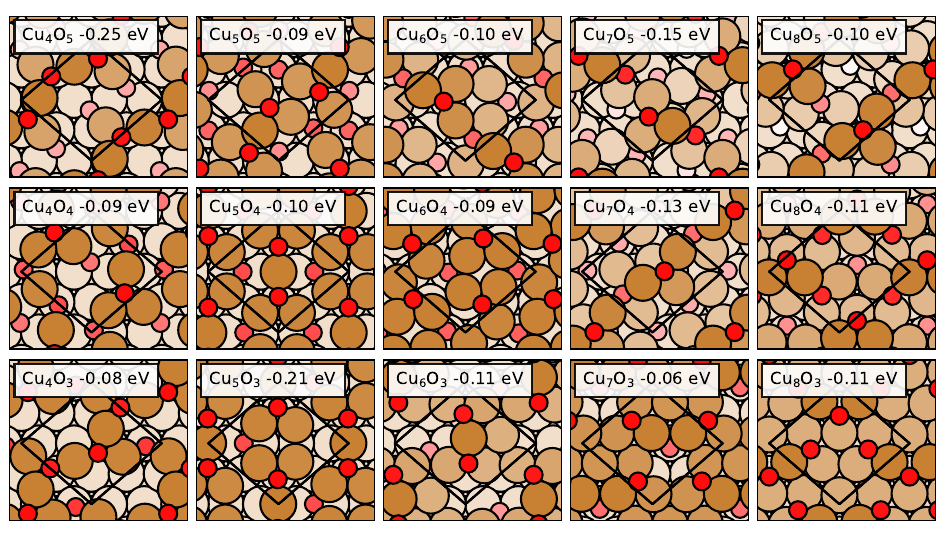}
    \caption{Minimum energy structures for each stoichiometry found with $\Delta$GP-CHGNet, the energies 
    reported in the insets are the difference between the lowest energy structure 
    found when employing a $\Delta$-model and those with just CHGNet, both calculated (including relaxations) with DFT.
    Small and large circles represent Cu and O atoms, respectively. The atoms
    are colored according to height above the slab with higher atoms being darker in color. The unit cell is shown 
    in black.}
    \label{fig:copper_structures}
\end{figure*}

\subsection{Global structure search}
\label{sec:gofee}

In the applications of the $\Delta$-learning augmented CHGNet model, we will study the stability of surfaces consisting of a fixed slab 
and an overlayer with variable stoichiometry. To find the optimal geometry for each stoichiometry 
we employ the GOFEE optimization algorithm as implemented in AGOX \cite{Bisbo2020, christiansenAtomisticGlobalOptimization2022}.
This algorithm iteratively explores the potential energy surface guided by a surrogate model by employing an acquisition
function that takes into account both the predicted energy and uncertainty of proposed structures. 
For our $\Delta$GP-CHGNet surrogate model, we use the uncertainty that can be calculated from the gaussian process model, 
whereas for searches with unmodified CHGNet no uncertainty is used. An outline of each iteration is as follows

\begin{enumerate}
    \item Create a number of structures.
    \item Locally optimize each structure in the lower-confidence bound of the current surrogate model (Either CHGNet or $\Delta$GP-CHGNet).
    \item Select the most promising candidate according to an acquisition function that takes the surrogate prediction into account.
    \item Perform a single-point DFT calculation for the selected candidate and store it in a database. 
    \item If using $\Delta$GP-CHGNet, update the surrogate model with the new data point.
\end{enumerate}

We run a number of such searches for each stoichiometry for a fixed number of iterations. 
We run searches with both CHGNet and $\Delta$GP-CHGNet, when using CHGNet there is no 
update step whereby the searches will be limited to exploring only structures that are 
local minima in CHGNet. With $\Delta$GP-CHGNet the $\Delta$GP-model is trained on single point DFT energies of 
the collected configurations, effectively establishing an active learning setting with focus on local minima structures.

This results in a number of configurations with total energies calculated at the DFT 
level for each stoichiometry. The DFT settings we employed are described in \label{sec:dft_appendix}.
To establish which stoichiometry/phase is stable under different conditions the Gibbs free energy is calculates as 

\begin{equation}
    \Delta G = E_{T} - E_{\mathrm{slab}} - \sum_Z n_Z (\Delta \mu_Z + \varepsilon_Z).
\end{equation}
Where $E_T$ is the total energy, $E_{\mathrm{slab}}$ is the total energy of the clean slab, 
$n_Z$ is the number of atoms with atomic number $Z$ in the overlayer and finally $\Delta \mu_Z$ and $\varepsilon_Z$ are 
the chemical potential and reference energy of atoms with atomic number $Z$.

\section{Results}

\subsection{Copper-oxide}

\begin{figure}[]
    \centering
    \includegraphics[width=0.45\textwidth]{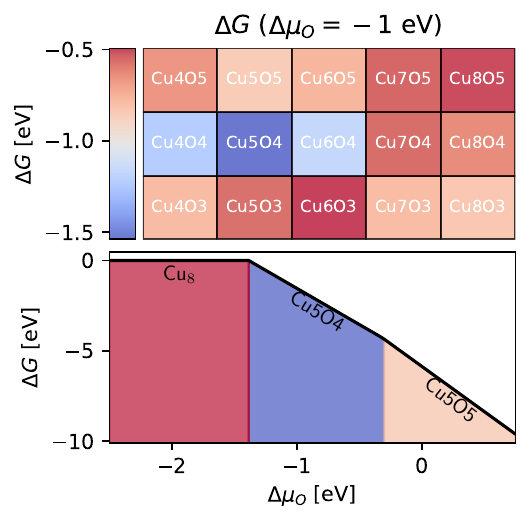}
    \caption{Top: Raster plot of the DFT-based Gibbs free energy at $\Delta \mu_O = -1$ eV. 
    Bottom: Phase diagram showing the most stable phase as a function of 
    the chemical potential of oxygen.}
    \label{fig:copper_phase}
\end{figure}

\begin{figure*}[ht!]
    \centering
    \includegraphics[width=0.75\textwidth]{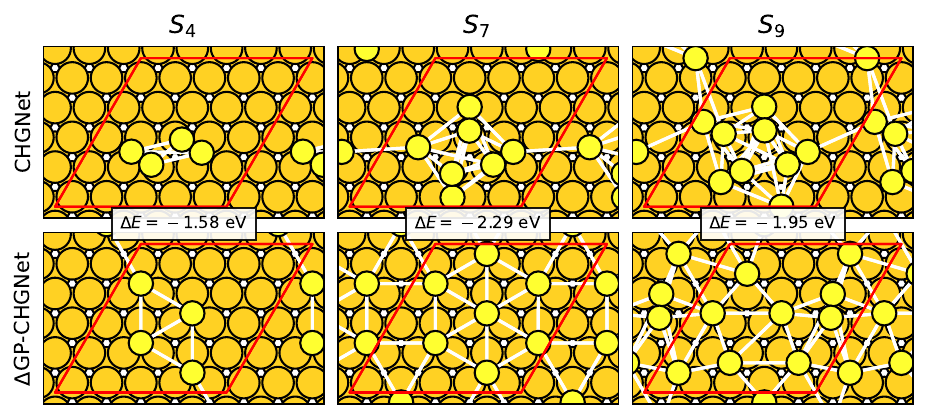}
    \caption{Minimum energy structures for three selected stoichiometries of sulfur on Au111 found with CHGNet (top) and $\Delta$GP-CHGnet (bottom). 
    The energies reported are calculated as $\Delta E = E_{\Delta \mathrm{GP-CHGNet}} - E_{\mathrm{CHGNet}}$
    where the subscripts indicate the method used to find the configurations but both are evaluated with DFT. The unit cell is shown in red.}
    \label{fig:gold_structures}
\end{figure*}

The first system we investigate is that of the "8" Cu oxide, which is
an ultra-thin oxide film on Cu(111). For this system, LEED shows
evidence of a periodic structure forming within a $c(8\times 4)$
surface cell, see Ref.\ \onlinecite{moritaniReconstructionCu111Induced2008} and references
therein.  A structural model containing 8 Cu and 4 O atoms was originally proposed
based on the experimental evidence in 2008 by Moritani \textit{et
  al.}\cite{moritaniReconstructionCu111Induced2008}. Recently, the model was revised based
on density functional theory calculations by Kim \textit{et
  al.}\cite{kimRecalibratingExperimentallyDerived2024}. However, no full phase diagram has been
constructed leaving this system open for discovery of new phases.

To supplement the previous studies, we have therefore considered
stoichiometries of the ultra-thin oxide film,
$\mathrm{Cu}_x\mathrm{O}_y$, with $ x = [4, 5, 6, 7, 8]$
and $y = [3, 4, 5]$. The reference energy for $\mathrm{Cu}$ is calculated as the difference 
in the total energy of a four and three-layer slab divided by the difference in the number of 
atoms. The reference energy for oxygen is half of the total energy of an $\mathrm{O}_2$ molecule 
from a spin-polarized calculation.

For each stoichiometry we employ the optimization algorithm described in Section \ref{sec:gofee} 
with CHGNet-v0.3.0 and with $\Delta$GP-CHGNet-v0.3.0. The lowest energy structures found by the 
searches with the $\Delta$GP model are reported in Fig. \ref{fig:copper_structures}. 
It is evident from this figure the structures found with the $\Delta$GP
are only slightly more stable than those found by just searching in CHGNet, for the majority of stoichiometries this can be attributed 
to minor local optimizations with no bearing on the configuration. 
This shows the remarkable ability of CHGNet to predict new, unpublished, structures for materials 
solely based on being trained on a wide variety of examples, the vast majority of which are 
bulk and not surfaces. 

The phase diagram for this system is shown in Figure
\ref{fig:copper_phase}.  Which shows that for a wide range of chemical
potentials for oxygen the preferred phase is Cu$_5$O$_4$. We note that
by considering a wide range of Cu$_x$O$_y$ stoichiometries, the
present search reveals a new structure that challenges the previous
assignment of Kim \textit{et al.} \cite{kimRecalibratingExperimentallyDerived2024}, which however is
also contained in Fig.\ \ref{fig:copper_phase} as the most stable
structure at that given stoichiometry, Cu$_8$O$_4$. While LEED-IV studies or
surface X-ray diffraction experiments would be required to
conclusively discriminate which structural model conforms best with
the actual Cu "8" oxide, it seems likely that the new Cu$_5$O$_4$ is
the correct structure as this aligns with the findings of an identical
structure for Ag(111)\cite{Schnadt2017ExperimentalOxygenAdsorption, mortensenAtomisticStructureLearning2020}
and a similar non-commensurate structure for Pd(111) \cite{lundgrenTwoDimensionalOxidePd1112002}.

\subsection{Sulfur adsorption on Au(111)}

\begin{figure}
    \centering
    \includegraphics[width=0.45\textwidth]{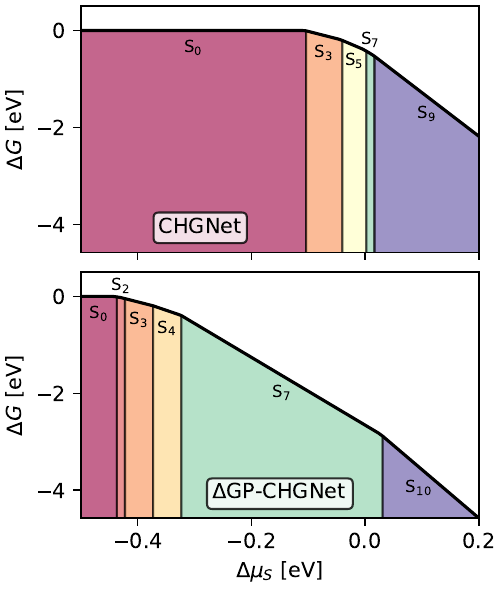}
    \caption{DFT-based Gibbs free energy phase diagrams as a function 
    of the chemical potential of sulfur for structures found 
    using out of the box CHGNet (top) and using $\Delta$GP-CHGnet (bottom). 
    The reference energy for sulfur is set as an eigth of the total energy of cyclooctasulfur.}
    \label{fig:gold_phases}
\end{figure}

\begin{figure*}
    \centering  
    \includegraphics[width=0.75\textwidth]{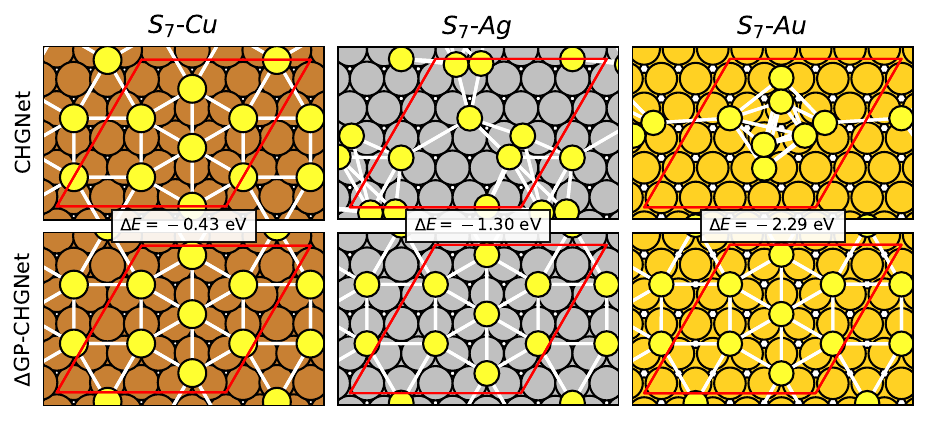}
    \caption{Sulfur overlayer global minimums identified with CHGNet (top) and $\Delta$-GP CHGNet (bottom). 
    The energy difference, with DFT, between the two structures is given. With $\Delta$-GP CHGNet 
    the same motif is found for all three metals, whereas without the correction dimers are formed 
    for silver and gold. CHGNet predicts the correct motif on copper, but it is shifted to sulfur on hcp hollow sites.}
    \label{fig:metal_sulfides}
\end{figure*}

The next system studied is that of sulfur adsorption on Au(111). For
this system, an adsorption phase having a $(5\times 5)$ super
cell has been observed with LEED, see Ref.\ \onlinecite{yuStructureAtomicSulfur2007}, and a surface structure with
7 sulfur atoms in a 'Rosette' was proposed as the underlying structure. The occurrence of this 
pattern is attributed to slightly repulsive interactions between adsorbed sulfur atoms. 
We investigate the efficacy of CHGNet and our proposed $\Delta$GP-CHGNet model for structural 
exploration of this system. \newline \newline
Again, we perform searches using only CHGNet and using $\Delta$GP-CHGNet, in Figure \ref{fig:gold_structures} 
the most stable configurations found with the number of sulfur atoms $N_S = [2, 7, 9]$ 
for both methodologies is shown. For this system, the differences in stability between structures identified purely with CHGNet and those 
found with $\Delta$GP-CHGNet are very significant. Out of the box CHGNet-v0.3.0 generally 
prefers sulfur dimers whereas $\Delta$GP-CHGNet learns that it is preferential for sulfur atoms 
to sit in separate hollow-fcc sites. This tendency may be observed in all three stoichoimetries depicted 
in Figure \ref{fig:gold_structures}. With four sulfur atoms, two dimers are arranged in 
a small cluster, with seven sulfur atoms three dimers and a lone sulfur atom form a cluster and 
finally for nine atoms four dimers and a lone sulfur form a cluster. In contrast, for all 
these structures, DFT prefers separated atoms forming extended overlayers -- only forming dimers 
when the number of sulfur atoms added to the surface is relatively high. \newline \newline
Given the large configurational discrepancies between the two search methodologies it is 
unsurprising that the Gibbs free energy diagrams differ extensively, as is shown in Figure 
\ref{fig:gold_phases}. For this system, if relying entirely on CHGNet no phases 
are predicted correctly except the trivial bare surface for which the chemical potential $\Delta \mu_S$
necessary for its occurrence is rather inaccurate. 

\subsection{Analysis}

So far, we have seen results for two systems. One where the universal potential 
CHGNet is able to predict a rather complex surface-oxide structure. However, 
for the other system the universal potential struggles significantly and consistently 
favors configurations with an erroneous bonding motif. For other metal-sulfur adsorption structures, such as 
S/Cu and S/Ag discrepancies between structures preferred in CHGNet and DFT  are also present but to 
a lesser extent. This is illustrated in Fig \ref{fig:metal_sulfides}.
In fact, for Cu(111)-(5$\times$5)-7S the correct motif is the global minimum 
of CHGNet, but it is shifted from sulfur sitting on fcc hollow sites to hcp hollow sites. 
For silver a cluster containing several dimers is the preferred structure according to 
CHGNet, whereas with DFT the same motif as for gold and copper is obtained. 

\begin{figure}
    \includegraphics[width=0.45\textwidth]{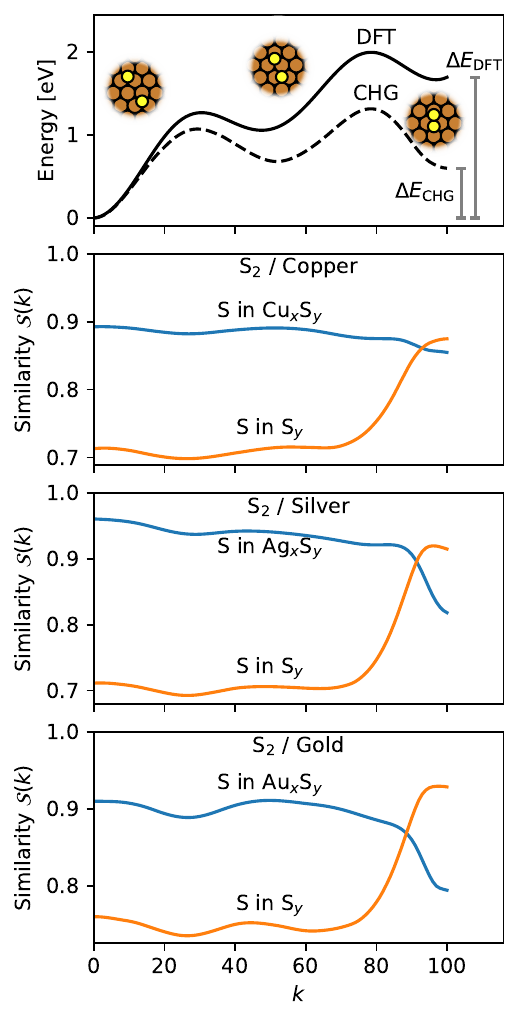}
    \caption{Energy profiles for combining two sulfur atoms on a copper surface. Average dot product similarity for sulfur atoms for a trajectory starting with separately adsorbed sulfur 
    atoms ending with a $S_2$-dimer. The blue lines shows the average similarity against sulfur atoms present in MPTrj 
    in structures of the form $X_xS_y$ where $X$ is copper, silver or gold. 
    The orange lines measures the similarity against sulfur atoms from MPtrj in structures containing only sulfur.}
    \label{fig:dimer_similarity}
\end{figure}

To facilitate an analysis of the origin of the poor CHGNet description
of S/Ag(111) and S/Au(111) systems, we have found it instructive to reduce the
complexity of the sulfur structure by studying systems with only two sulfur atoms present in the cell. For
all three metals Cu, Ag, and Au it can be found with DFT that
configurations in which the two sulfur atoms are separated are more
stable than configurations in which they form a dimer on the
surface. A full pathway from two separated sulfur atoms to a sulfur
dimer can be calculated, see Fig.\ \ref{fig:dimer_similarity} for an
example for the Cu(111) surface.

\begin{table}
    \centering
    \begin{tabular}{| c | c | c |}
        \hline
        Element  &  Configurations  &  Environments  \\
        \hline
        Cu + O& 647 & 3268 \\
        Cu + S& 548 & 5154 \\
        Ag + S& 374 & 828 \\
        Au + S& 17 & 42 \\
        S only& 636 & 20450 \\ 
        \hline
    \end{tabular}
    \caption{Number of configurations in MPTrj of the form M$_x$S$_y$ for different metals and the 
    number of configurations containing only sulfur along with the total number of sulfur environments 
    for each type of configuration. In addition, the table lists the number of Cu$_x$O$_y$ configurations and the number of 
    oxygen environments in those in MPtrj.}
    \label{table:sulfur_counts}
\end{table}


In an effort to investigate the reasons for this difference in CHGNet's ability to 
predict physically realistic results we wish to understand which parts of the Materials Project training 
data are likely to have influenced the predictions for each system. For this reason, 
we compute the CHGNet features for every sulfur for various metal-sulfides present 
in the MPtrj dataset alongside the features of sulfur atoms of the $S_2$ system for trajectories starting 
with the separated sulfur atoms and ending with a sulfur dimer. Such features may be compared 
using a similarity metric, such as a normalized dot product. For each configuration of the 
$S_2$-trajectories we compute
\begin{equation}
    \mathcal{S}(k) = \frac{1}{2N} \sum_i^N \sum_j^2 \frac{\vec{v}_i \cdot \vec{w}_j(k)}{|\vec{v}_i||\vec{w}_j(k)|}
\end{equation}
Where $\vec{v}_i$ is the CHGNet representation of a sulfur-atom in the 
MPtrj dataset, either from configurations involving both sulfur and metal atoms or from configurations
involving only sulfur atoms. $\vec{w}_j(k)$ is the representation  of the
$j$'th sulfur atom for configuration $k$ along the trajectory. 
The result of this is shown in Fig. \ref{fig:dimer_similarity}.
On copper the similarity towards sulfur atoms in environments containing both Cu and S remains largely constant along the pathway, while the similarity involving only S increases as the dimer is formed.
In contrast, on both silver and gold the similarity towards 
sulfur atoms in environments containing the metal decreases sharply at the same time as the similarity
towards sulfur in environments containing only sulfur increases. That is, 
for sulfur on silver and gold CHGNet relies too heavily on information gathered from pure sulfur 
configurations where dimers are very stable. This may 
explain why CHGNet can identify that on copper the dimer is not preferential, while on 
silver and gold it is unable to do so -- which then leads to erroneous configurations when 
additional sulfur atoms are introduced. 

Further evidence for this explanation is found from the number of sulfur environments originating 
from configurations of different types present in the MPtrj training dataset, presented in 
Table \ref{table:sulfur_counts}. For Cu$_x$S$_y$ the number of configurations is comparable 
to the number of configurations of pure sulfur, whereas for silver and especially for gold the 
number of configurations and environments is substantially less than those of pure sulfur.
It stands to reason, that this imbalance has at least partially induced the overreliance on
sulfur environments from configurations containing only sulfur. 

\begin{figure}
    \includegraphics[width=0.45\textwidth]{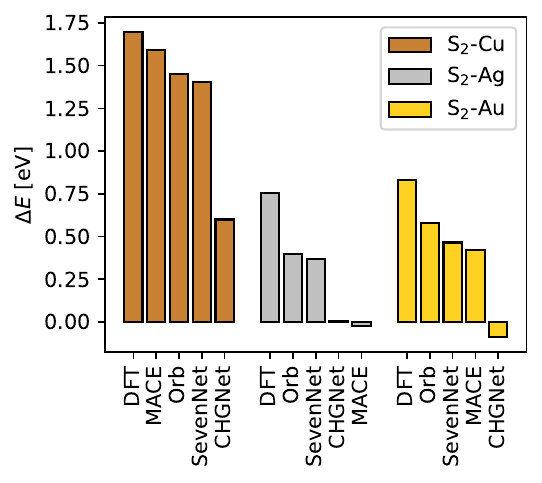}
    \caption{Relative stability of sulfur-dimer compared to two separated sulfur atoms on copper, silver and gold. 
    Relative stabilities have been calculated using DFT, CHGNet-0.3.0, MACE-MP0, SevenNet-0 and ORB-v2-only-MPtrj -- 
    which all are trained on the MPtrj dataset. Positive values indicate that the dimer is less stable 
    than the two separated atoms. A visual representation of these numbers are depicted in Fig. \ref{fig:dimer_similarity}(a)}
    \label{fig:compare_uni}
\end{figure}

CHGNet is not the only universal potential trained on MPtrj. In Figure \ref{fig:compare_uni} we 
show the relative stability of $S_2$ dimers compared to two separated $S$ atoms on the 
three different metals calculated with DFT and four machine learning potentials. 
All of them overestimate the stability of the dimer, but by various amounts and as 
with CHGNet the errors are generally smallest when the metal is copper. 

\section{Conclusion}

We have investigated the use of a universal potential, CHGNet, for the task of global mimimum energy structure prediction, 
specifically for two systems an oxygen induced surface reconstruction of copper and 
sulfur overlayers on group 11 metals. Further, we introduce a $\Delta$-learning method 
on top of the universal potential in order to perform global optimization searches 
with both the out-of-the-box CHGNet and this adapted $\Delta$-GP-CHGNet model. 

For the "8" Cu surface oxide we find a new global minimum structure and 
show that this is a discovery that as-is CHGNet is capable of supporting. Whereas we for the sulfur overlayer 
systems find that CHGNet has flawed understanding that leads to false predictions for the 
global minimum energy structures. Analyzing the behavior of CHGNet for the S$_2$ dimer in various metallic embeddings, we traced the origin of the false predictions to CHGNet overestimating the stability 
of the sulfur-dimer. By inspecting the MPtrj dataset using the representations 
CHGNet has learned during its training on this dataset, we find that a possible cause for 
CHGNets mistaken understanding of adsorbed sulfur dimers is an overreliance on the parts 
of the training data that only involve sulfur. 

Universal potentials provide the materials science community with the opportunity to 
investigate more and larger systems and allows realistic materials modelling to be done 
with much fewer computational resources. However, they do not come with guarantees and 
may, as we have shown, be severely mistaken. The method we have presented is an effective
way of correcting such mistakes with minimal overhead by relying on the 
representations learned by the universal potential. This requires the descriptors to be extractable while remaining connected to the computational graph. 
Therefore, we encourage developers of such models to prioritize making their code more extensible.
Users should be wary and make sure to test the correctness of any such potential, at least for a scaled-down version of 
their system of interest. Our efforts are one way of adding some explainability to 
the predictions of a universal potential, further work may involve the application of 
other model interpretability methods such as TracInCP for the identification of 
influential training examples \cite{pruthiEstimatingTrainingData2020}. 

\section{Data availability}
The datasets generated and analyzed in this manuscript are openly available at https://doi.org/10.5281/zenodo.15090225. 

\section{Code availability}
The code used in this manuscript is openly available on GitLab at https://gitlab.com/agox/agox-chg.

\section{Acknowledgements}
We acknowledge support from VILLUM FONDEN through Investigator grant, project 
no. 16562, and by the Danish National Research Foundation through the Center of 
Excellence “InterCat” (Grant agreement no: DNRF150).

\bibliographystyle{apsrev4-1}
\bibliography{references}

\appendix
\section{Appendix: DFT Settings}
\label{sec:dft_appendix}
All DFT calculations have been performed with the GPAW code \cite{LarsenASE2017, mortensenGPAWOpenPython2024}. For copper-oxide a plane-wave cutoff of 400 eV was used, a (4$\times$4) Monkhorst-Pack k-point grid, with the PBE 
exchange-correlation functional. For the metal-sulfides we employed a cutoff of 520 eV and Monkhorst Pack grids with a density of $3.5$ points/Å$^{-1}$ again with the PBE 
functional.

\end{document}